\documentclass{bmcart}

\usepackage{amsthm,amsmath}
\usepackage{graphicx}
\usepackage{hyperref}
\usepackage[utf8]{inputenc} 
\usepackage{lmodern}
\usepackage{color}
\usepackage{booktabs}

\startlocaldefs
\graphicspath{{images/}}

\newcommand{\eref}[1]{Eq.~(\ref{#1})}
\newcommand{\esref}[2]{Eqs.~(\ref{#1})-(\ref{#2})}
\newcommand{\sref}[1]{Section~\ref{#1}}
\newcommand{\fref}[1]{Fig.~\ref{#1}}
\newcommand{\tref}[1]{Table~\ref{#1}}
\newcommand{\mcN}{\mathcal{N}}

\endlocaldefs

\begin{document}

\begin{frontmatter}

\begin{fmbox}
\dochead{Research}


\title{Universal temporal features of rankings in competitive sports and games}


\author[
   addressref={ciencias},
   email={jamafcc@ciencias.unam.mx}
]{\inits{JA}\fnm{Jos{\'e} A.} \snm{Morales}}
\author[
   addressref={ciencias},
   email={sergio_sanchez@ciencias.unam.mx}
]{\inits{S}\fnm{Sergio} \snm{S{\'a}nchez}}
\author[
   addressref={fisica},
   email={jfv@fisica.unam.mx}
]{\inits{J}\fnm{Jorge} \snm{Flores}}
\author[
   addressref={fisica,c3},
   corref={fisica},
   email={carlosp@fisica.unam.mx}
]{\inits{C}\fnm{Carlos} \snm{Pineda}}
\author[
   addressref={iimas,c3,mit,neu,itmo},
   email={cgg@unam.mx}
]{\inits{C}\fnm{Carlos} \snm{Gershenson}}
\author[
   addressref={fisica},
   email={cocho@fisica.unam.mx}
]{\inits{G}\fnm{Germinal} \snm{Cocho}}
\author[
   addressref={ciencias},
   email={zepeda@fisica.unam.mx}
]{\inits{J}\fnm{Jer{\'o}nimo} \snm{Zizumbo}}
\author[
   addressref={cide,aalto},
   email={gerardo.iniguez@cide.edu}
]{\inits{G}\fnm{Gerardo} \snm{I{\~n}iguez}}


\address[id=ciencias]{
  \orgname{Facultad de Ciencias, Universidad Nacional Aut{\'o}noma de M{\'e}xico},
  \postcode{01000}
  \city{M{\'e}xico D.F.},
  \cny{Mexico}
}
\address[id=fisica]{
  \orgname{Instituto de F{\'i}sica, Universidad Nacional Aut{\'o}noma de M{\'e}xico},
  \postcode{01000}
  \city{M{\'e}xico D.F.},
  \cny{Mexico}
}
\address[id=iimas]{
  \orgname{Instituto de Investigaciones en Matem{\'a}ticas Aplicadas y en Sistemas, Universidad Nacional Aut{\'o}noma de M{\'e}xico},
  \postcode{01000}
  \city{M{\'e}xico D.F.},
  \cny{Mexico}
}
\address[id=c3]{
  \orgname{Centro de Ciencias de la Complejidad, Universidad Nacional Aut{\'o}noma de M{\'e}xico},
  \postcode{04510}
  \city{M{\'e}xico D.F.},
  \cny{Mexico}
}
\address[id=mit]{
  \orgname{SENSEable City Lab, Massachusetts Institute of Technology},
  \postcode{02139}
  \city{Cambridge MA},
  \cny{USA}
}
\address[id=neu]{
  \orgname{MoBS Lab, Network Science Institute, Northeastern University},
  \postcode{02115}
  \city{Boston MA},
  \cny{USA}
}
\address[id=itmo]{
  \orgname{ITMO University},
  \postcode{199034}
  \city{St. Petersburg},
  \cny{Russian Federation}
}
\address[id=cide]{
  \orgname{Centro de Investigaci{\'o}n y Docencia Econ{\'o}micas, Consejo Nacional de Ciencia y Tecnolog{\'i}a},
  \postcode{01210}
  \city{M{\'e}xico D.F.},
  \cny{Mexico}
}
\address[id=aalto]{
  \orgname{Department of Computer Science, Aalto University School of Science},
  \postcode{FI-00076}
  \city{AALTO},
  \cny{Finland}
}



\end{fmbox}

\begin{abstractbox}

\begin{abstract} 
Many complex phenomena, from the selection of traits in biological systems to hierarchy formation in social and economic entities, show signs of competition and heterogeneous performance in the temporal evolution of their components, which may eventually lead to stratified structures such as the wealth distribution worldwide. However, it is still unclear whether the road to hierarchical complexity is determined by the particularities of each phenomena, or if there are universal mechanisms of stratification common to many systems. Human sports and games, with their (varied but simplified) rules of competition and measures of performance, serve as an ideal test bed to look for universal features of hierarchy formation. With this goal in mind, we analyse here the behaviour of players and team rankings over time for several sports and games.  Even though, for a given time, the distribution of performance ranks varies across activities, we find statistical regularities in the dynamics of ranks. Specifically the rank diversity, a measure of the number of elements occupying a given rank over a length of time, has the same functional form in sports and games as in languages, another system where competition is determined by the use or disuse of grammatical structures. Our results support the notion that hierarchical phenomena may be driven by the same underlying mechanisms of rank formation, regardless of the nature of their components. Moreover, such regularities can in principle be used to predict lifetimes of rank occupancy, thus increasing our ability to forecast stratification in the presence of competition.
\end{abstract}



\end{abstractbox}

\end{frontmatter}

\section{Introduction} 
\label{sec:intro}

Sports and games can be described as complex systems due to the myriad of
factors influencing the dynamics of competition and performance in them,
including networked interactions among players, human and environmental
heterogeneities, and other traits at the individual and group
levels~\cite{duch2010quantifying,ben2007most,merritt2013environmental,merritt2013social}.
In particular, player performance is influenced by a variety of causes:
Economical, political and geographical conditions determine the ranking of a
single player and thus may be used for predicting performance. Moreover, the
(relatively) simple rules of competition and measures of performance associated
with sports and games allow us to explore basic mechanisms of interaction
leading to hierarchy formation that may be common to many systems driven by
competition, not only leisure activities but other social, biological and
economic systems. With this goal in mind, the availability of a large corpus of
data related to sports, teams, and players allows researchers to perform
multiple statistical analyses, in particular with respect to the structure and
dynamics of player rankings~\cite{albert2005anthology,radicchi2011best,yucesoy2016untangling}. 

Data availability has made it possible not only to study the distribution of
scores determining rankings, but also its time
evolution~\cite{merritt2014scoring}. In a recent paper, Deng et al. present a
statistical analysis of 12 sports showing a universal scaling in rankings,
despite the fact that the sports considered have very different ranking
systems~\cite{1367-2630-14-9-093038}. Here, we focus on the temporal
trajectories of player and team performances, meaning the evolution of rank,
with the objective of finding statistical regularities that indicate how
competition shapes hierarchies of players and teams. In general, rankings are
affected in time by events as apparently insignificant as a bad breakfast prior
to an important event, or the weather during a
competition~\cite{klaassen2001points}. Since these factors are inherently
present for all activities, we would expect the evolution of rank to have
universal features across sports and games. 

We propose to quantify such evolution by means of a recently introduced
measure, the {\it rank diversity}. With the help of the Google $n$-gram
dataset~\cite{Michel14012011}, rank diversity has been used before to study how
vocabulary changes in time~\cite{cocho2015rank}. That work shows that rank
diversity has the same functional form for all languages studied, and is able
to discriminate the size of the core of each language. Thus, we concentrate on
the temporal features of rank distributions corresponding to several sports and
games with different ranking schemes. We consider data where an appropriate
time resolution is available, and limit the analysis to six activities only:
tennis, chess, golf, poker and football (national teams and clubs). We find
that all rank diversities have the same functional form as languages, despite
having very different rank frequency distributions. Finally, we introduce a
random walker model that, tuned by the parameter values of each dataset,
reproduces qualitatively the diversity of all sports and games considered. 

The article is organized as follows. In section \sref{sec:rankData} we describe
the datasets used. We then analyse ranking distributions in
\sref{sec:sportsRank} and compare them with several models. Finally, in
\sref{sec:rankDiv} we study the rank diversity for each sport and compare it
with the random walker model. The main conclusions of our analysis are stated
in \sref{sec:conc}. 

\section{Ranking data} 
\label{sec:rankData}

We use ranking data on players and teams from six sports and games: a) Tennis players (male), ranked by the Association of Tennis Professionals (ATP)~\cite{atpWorldTour}; b) Chess players (male), ranked by the F\'ed\'eration Internationale des \'Echecs (FIDE)~\cite{worldChessFed}; c) Golf players, ranked by the Official World Golf Ranking (OWGR)~\cite{officialWorldGolf}; d) Poker players, ranked by the Global Poker Index (GPI)~\cite{globalPokerInd}; e) Football teams, ranked by the Football Club World Ranking (FCWR)~\cite{footballClubWorld}; and f) national football teams, ranked by the F\'ed{\'eration Internationale de Football Association (FIFA)~\cite{fifa}.

The ranking procedure varies among sports. In ATP, for example, tennis players are ordered according to the number of points they have up to the date of publication of the ranking. The number of points depends on the tournaments players have participated in (and how well they have performed), but not all tournaments are taken into account. FIDE uses the Elo system~\cite{elo1978rating} to rank players, which considers the number of matches, their results, and the opponent ranking. The FIFA ranking takes into account official
matches between countries. The number of points depends on the confederation and classification of each team, as well as the importance and result of the match. \tref{tab:ranking_data} summarizes the main properties of the ranking data used in this study, including the temporal resolution of rankings. In order to have a homogeneous ranking resolution for each sport, we disregard some data for the ATP, FIDE, and GPI datasets. In poker, for example, due to the characteristics of some tournaments, rankings for periods shorter than a week may exist, yet we only use weekly data.

\section{Comparison with ranking models} 
\label{sec:sportsRank}

Player or team performance is usually measured by a {\it score} that varies with time. This score results in a time-dependent {\it rank} with rather complex behaviour, as we will explain below. We first focus on the distribution of scores versus ranks (i.e. a {\it rank distribution}) for a given time. Particularly, we are interested in seeing if this distribution can be reproduced by a single ranking model for all sports and games considered. We select five ranking models to fit the data, four of which are particular
cases of
\begin{equation}
f(k)=
\mcN
\frac{(N+1-k)^{d}\exp(-bk)}{k^{a}},
\label{eq:m:general}
\end{equation}
where $f$ is the score associated with rank $k$, $a$ is an exponent that dominates most of the curve, $b$ an exponent controlling its exponential decay, and $d$ an algebraic decay that regulates a sharp drop of the curve for its last elements. Finally, $N$ is the total number of elements (i.e. players or teams) in the system, and $\mcN$ is a normalization constant.

The first four models are
\begin{equation}
m_1(k)  \propto \frac{1}{k^a}, \quad
m_2(k)  \propto \frac{\exp(-bk)}{k^a}, \quad
m_3(k)  \propto \frac{(N+1-k)^{d}}{k^a}, \quad
m_4(k) = f(k), 
\label{eq:models:one:to:four} 
\end{equation}
whereas the fifth model is a double Zipf law~\cite{PhysRevX.3.021006},
\begin{equation}
m_5 (k)=\mathcal{N} \begin{cases}
             \frac{1}{k^a} &   k\leq k_c \\
             \frac{k_c^{a'-a}}{k^{a'}} &  k>k_c  
             \end{cases},
\label{eq:double:zipf}
\end{equation}
with $a'$ an alternative exponent that regulates the behaviour of the curve
after a critical rank $k_c$.
Model $m_1$ is
obtained by setting $d=b=0$ in \eref{eq:m:general}, and has been considered in
a vast amount of studies, both in the realm of
sports~\cite{doi:10.1080/026404199365777,AlvarezRamirez2006509} and in other
studies of ranking behaviour~\cite{1367-2630-15-4-043021}, including the famous
Zipf's law of languages where the particular case $a=1$ has drawn a lot of
attention (see, e.g.,~\cite{1367-2630-13-4-043004} and references therein). The
Gamma ($m_2$) and Beta ($m_3$) distributions have been useful in many
disciplines for decades; a quick look at their Wikipedia entries provides
numerous examples~\cite{wiki:gamma, wiki:beta}.  Model $m_4$, being a more
general expression than the previous ones, tends to provide a better fit at the
expense of more parameters, and will serve as benchmark for the comparison
between the rest of the models. Finally, model $m_5$ in \eref{eq:double:zipf}
has been used with success in several
contexts~\cite{PhysRevX.3.021006,1538-4357-640-1-L5}, prompting us to test it
in the area of sports and games.

The results of the fitting process between data and
\esref{eq:m:general}{eq:double:zipf} are shown in \fref{fig:rank}, while
\tref{tab:fitting_rank_data} summarises the parameter values obtained. Data
corresponds to a single time snapshot for all sports and games: Sept 25 2014 (ATP); Sept 2014
(FIDE); Mar 18 2015 (GPI); Apr 19, 2015 (OWGR); Dec 29 2014 (FCWR); and Dec 18
2014 (FIFA). Both models and sports show variation in their goodness of fit.
From \fref{fig:rank} it is clear that Zipf's law ($m_1$) is not adequate. On
the other hand, the Gamma distribution ($m_2$) fits some datasets rather well,
particularly those that do not show an abrupt fall of score as a function of
rank (ATP and GPI). Datasets with an abrupt decay of frequency are well fitted
by the Beta distribution ($m_3$) (FIDE, OWGR, and FIFA). FCWR is an
intermediate case where both functions seem to capture global behaviour
accurately, and thus the fit is considerably better for a combination of both
models, i.e. $m_4$. We also see that the double Zipf law ($m_5$) is a good fit
for GPI, as seen from \tref{table:goodnes:of:fits}. 

In order to objectively compare goodness of fit between models, we consider
several measures: The least-squares parameter $R^2$, the maximum deviation
between theory and observation $D$, and the Kolmogorov index
$p$~\cite{kolmogorov1933sulla}.
$R^2$ is calculated as the 2-norm
with respect to the data, that is, $R^2 = \sum_k [m_{i}(k)-y_k)]^2$ for a given
model fit $m_{i}(k)$, $i = 1, \ldots, 5$, and data $y_{k}$. The closer $R^2$ is
to one, the better the fit is. 
To calculate $D$, we consider the cumulative of both the proposed 
distribution [here, $m_i(k)$] and the dataset with 
$N$ data points. 
For a given model $m_i(k)$, the cumulative is simply $M_i(k) = \int_{-\infty}^k
m_i(\kappa) d \kappa$, whereas for a dataset, $M_{\rm data}(k) =(1/N) \sum_i
\theta(k-k_i)$, with $\theta$ a step function~\cite{e12071743}.
We then define $D$ as the maximum vertical
difference between the two curves, that is, $D = \max_k|M_i(k) - M_{\rm
data}(k)|$.  The Kolmogorov index $p$ is calculated as follows: Consider the
value of $D$ for a given dataset.  We then generate the same number of points
$N$ using the proposed model, say $m_i$, and repeat the procedure to obtain a
new $\tilde D$.  We repeat several times to obtain a set of $\tilde D$ values
(in our case, 2500).
The index $p$ is the fraction of
synthetic datasets that have larger $D$ than the original dataset
analysed. 
The measure $p$ allows us to consider that a small dataset will have
some noise due to poor statistics. Thus, if a model is consistent with 
a dataset, but we have poor statistics, we might still have a good 
(large) $p$. 
Usually, a ``good'' fit is required to have
$p>0.1$, see e.g.~\cite{clauset2009power}.

\tref{table:goodnes:of:fits} summarises the values of $R^2$, $D$, and $p$ found
for the six sports and five models $m_i$. None of the models are a good fit for
all sports, although $m_4$ is the most appropriate in terms of $R^2$ for most
sports. The exception is OWGR, where $m_5$ fits slightly better. However, in
three cases (FIDE, GPI, and FCWR) we have $p=0$ for model $m_4$, and no model
fits well, meaning that the theoretical distribution is not followed by the
data. For the other three sports (OWGR, FIFA, and ATP), 
$m_2$, $m_3$, and $m_4$, respectively, are the best fit. At least with the data and
models considered, we find no signs of universality. We stress again that
Zipf's law ($m_1$) is the worst fit among all considered. It is interesting to
notice that $R^2$ and $p$ lead to different criteria of what is a `good' or
`bad' model. This is due to both the amount of available data and the number of
parameters in the model. The larger the data, the easier it is to distinguish
an appropriate model from a good (but not accurate enough) approximation. On
the other hand, the more parameters the model has, the easier it is to fit any
data. Both of these aspects are taken into account in the definition of $p$,
but not in $R^2$.

\section{Rank diversity in sports} 
\label{sec:rankDiv}
In this paper we contribute to the analysis of sports ranking by computing the
rank diversity, a measure of the number of elements occupying a given rank over
a length of time. It appears that rank diversity has the same functional form,
not only for sports but also for other complex systems, such as countries
classified by their economic complexity, the 500 leading enterprises ranked by
the {\it Fortune} magazine, or a set of millions of words in six Indo-European
languages~\cite{cocho2015rank}.

The rank diversity $d(k)$ is defined as the number of distinct elements in a
complex system that occupy the rank $k$ at some point during a given length of
time.  In other words, we choose to focus on the time dependence of ranks,
rather than on a static distribution such as $f(k)$. An example of the change
of ranks in time for the sports and games studied here can be seen in
\fref{fig:spaghettis}. These so-called ``spaghetti'' curves show how
elements---individuals
or teams---change their rank in time. The rank diversity $d(k)$ is simply the
normalized number of different elements (curves) that spend at least one time
interval at a given rank $k$. The rank diversity for the various sports and games is
shown in \fref{fig:diversity}.

Studying $d(k)$ for six Indo-European languages~\cite{cocho2015rank}, we found
that the observed rank diversity closely follows the cumulative of a Gaussian (i.e. a sigmoid)
\begin{equation}
\Phi_{\mu,\sigma}(\log k)=\frac{\max_i  d(k_i)}{\sigma\sqrt{2\pi}}
   \int_{-\infty}^{\log k}\exp\left(-\frac{(y-\mu)^2}{2\sigma^2}\right) {\rm d} y.
   \label{eq:Phi}
\end{equation}
The mean value $\mu$ is set as the smallest $k_0$ for which 
$d(k_0)=\frac{\max_{ i } d(k_i) }{2} $, 
while the width $\sigma$ is fitted and gives the scale for which $d(k)$
gets close to its extreme values. If $k_{\pm}$ are given by
$\log_{10}k_{\pm}=\mu\pm2\sigma$,
the bulk of the changes in the values of diversity lies between $k_{-}$ and
$k_{+}$. In \fref{fig:diversity} we show the fit $\Phi$ for all
sports and games considered here ($R^2$ values for the $\Phi$ curves are shown there as well).
We do not consider neither $D$ nor $p$, since
these measures are only meaningful for distributions, which $d(k)$ is not.
To compare different rank diversity curves, their rank can be normalised to
$\frac{\log(k)-\mu}{\sigma}$, as shown in \fref{fig:normalized}. Since all the
cases considered can be fitted with the sigmoid curve of \eref{eq:Phi}, we argue that the rank diversity of sports seems to have a universal shape.
\subsection{A random walk model} 

Previously we have proposed a simple model to describe the evolution of rank
diversity~\cite{cocho2015rank}. We call this model a scale-invariant
random Gaussian walk, since a member with rank $k_{t}$, at the discrete time
$t$, is converted to rank $k_{t+1}$ according to the following procedure: We
define an auxiliary variable $s_{t+1}$ at time $t+1$ by the relation
\begin{equation}
s_{t+1}=k_{t}+G(k_t\hat{\sigma}),
\end{equation}
where $G(\tilde{\sigma})$ is a Gaussian-distributed random number with standard
deviation $\tilde{\sigma}$ and mean 0. This means that the random variable
$s_{t+1}$ has a width distribution proportional to $k_t$, and will thus, for
small $k_t$, have small changes as well. Once the values of $s_{t+1}$ for all
members are obtained, we order them according to their magnitude. This new
order gives new rankings, i.e. the $k$ values at time $t+1$. The only parameter
left in the model is the relative width $\hat\sigma$, which we deduce from the
empirical data. In \fref{fig:diversity:simulated} we show the diversity for
systems with the same number of elements as those of \fref{fig:diversity}, but
generated with the random model. It can be seen that these two sets of plots
are qualitatively similar. 

Notice that members with very low ranks will change very slowly or not at all,
while those with higher $k$ have a larger rank variation in time, as reflected
by $d(k)$. This intuition is clear from recent experience in sports: Federer, Nadal and Djokovic have
been the only number one tennis players in the twenty first century. The same holds for football clubs:
Barcelona, Bayern M\"unchen, Manchester United
and Real Madrid have been the clubs with lower ranks for many years. In other
words, members with small $k$ have a small rank diversity.

\section{Discussion and conclusions} 
\label{sec:conc}

Competition and heterogeneous performance are characteristic of the elements of
many complex systems in biological, social and economic settings. Despite the
fact that these systems show a large variation in the definitions of their
constituents and the relevant interactions between them, it remains to be seen
whether the emergence of hierarchical structure is mostly determined by the
particularities of each phenomenon, or if there are mechanisms of
stratification common to the temporal evolution of many systems.  We have
explored this notion by considering a set of relatively controlled and
simplified systems driven by competition: Human sports and games,
where the rules of engagement and measures of performance are well defined, in
contrast to, say, the ranking of physicists (the question of whom is the `best'
physicist would
have an ambiguous answer, to say the least). This allows us to characterise
the emergence of hierarchical heterogeneity by comparing the temporal features
of rankings of individuals and teams across activities in a clear way.
Explicitly, we analysed the statistical properties of rank distributions in six
sports and games, each with different number of members and rules for
calculating scores (and, therefore, ranks). By comparing rank distributions
with several ranking models, we find that the Zipf law
(model $m_1$) does not provide a suitable fit for the empirical data. Even if the more generic ranking model $m_4$ (a combination of the Gamma and Beta
distributions) tends to offer good fits, it is not always the best. This implies that rank distributions are not universal, at least with the models considered.

Furthermore, we studied the temporal features of rankings explicitly by calculating the rank diversity $d(k)$, a measure of the number of individuals or teams occupying a given rank over a length of time. Regardless of differences in the rank distribution across activities, $d(k)$ has the same sigmoid-like functional form, even for relatively small systems like FIFA (with only 150 elements per time slice). Coupled to the fact that a sigmoid rank diversity has also been found in the way vocabulary changes in time~\cite{cocho2015rank}, our results suggest that the emergence of hierarchical complexity -- as measured by $d(k)$ -- may have traits common to many systems.

A natural direction to follow in the near future is to study the behaviour of rank diversity in other competitive phenomena beyond sports and language, such as physical, social and economic processes of stratification. If indeed a certain universality in the temporal features of rankings is present in other complex settings, it would indicate that hierarchical phenomena may be driven by the same underlying mechanisms of rank formation, regardless of the nature of their components. Potentially, we may exploit such regularities to predict lifetimes of rank occupancy, thus increasing our ability to forecast stratification in the presence of competition.


\begin{backmatter}

\section*{Acknowledgements}
Financial support from CONACyT under projects 212802, 221341, and 
UNAM-PAPIIT IN111015 is acknowledged. 

\section*{Open access}
This article is distributed under the terms of the Creative Commons Attribution 4.0 International License (http://creativecommons.org/licenses/by/4.0/), which permits unrestricted use, distribution, and reproduction in any medium, provided you give appropriate credit to the original author(s) and the source, provide a link to the Creative Commons license, and indicate if changes were made.

\section*{Electronic Supplementary Material}
The datasets supporting the conclusions of this article are included as additional files. Below is the link to the electronic supplementary material.

\section*{Competing interests}
The authors declare that they have no competing interests.

\section*{Author's contributions}
All authors made substantial contributions to the conception and design of 
the paper and interpretation of data. They were all involved in drafting the
manuscript by contributing with relevant content. JAM and SS also contributed
with the acquisition and analysis of data. All authors read and approved the
final manuscript.


\bibliographystyle{bmc-mathphys} 
\bibliography{references} 
\section*{Figures} 

\begin{figure} 
\includegraphics[width=0.9\textwidth]{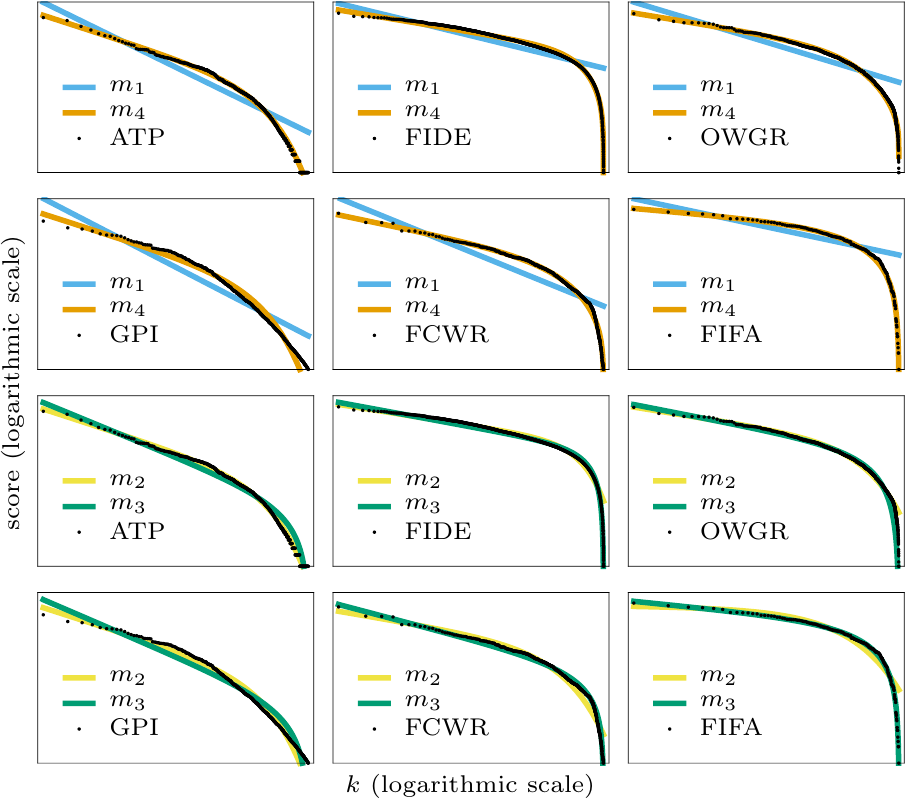}
\caption{\csentence{Comparison of ranking data with several models.}
Plot showing the rank distributions (score versus rank $k$) for all sports
considered in this study, at one time slice, as well as their
fits with the four models
in \esref{eq:m:general}{eq:models:one:to:four}. In general, Zipf's
law ($m_1$) does not accurately reproduce the ranking of any sport or game
reported here. Both the Gamma ($m_2$) and Beta ($m_3$) distributions seem to be
appropriate in some cases. Overall, all datasets are well fitted by a
combination of both functions, model $m_4$. For
the sake of clarity, $m_5$ and tick labels are not shown.
}
\label{fig:rank}
\end{figure} 

\begin{figure} 
\includegraphics[width=\textwidth]{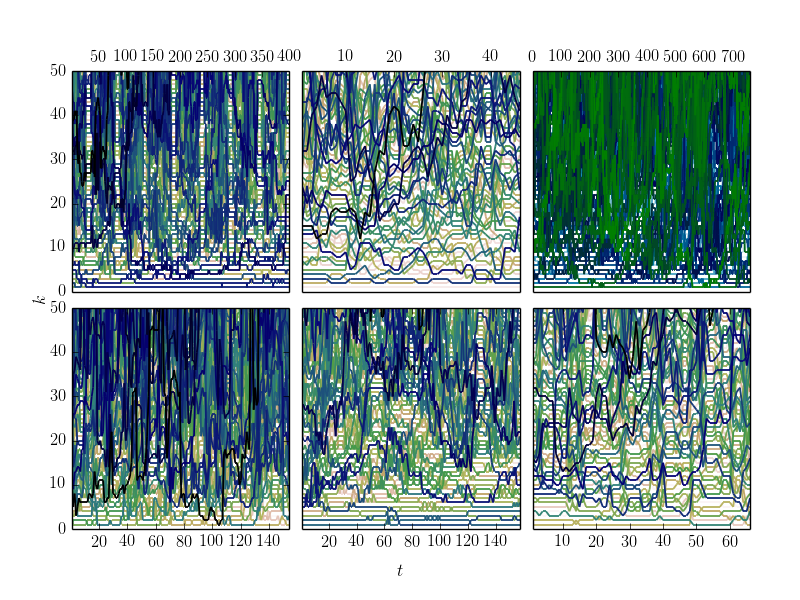}
\caption{\csentence{Temporal evolution of player and team rankings.} 
Plot showing the change in rank $k$ across time $t$ of all players/teams in
each sport and game considered in this study: ATP, FIDE, OWGR, FIFA, FCWR, and GPI (clockwise, starting from upper left
corner). Only the first 50 ranks at all available time slices are shown.
Notice that lower ranks tend to change less than higher ones, even when
different sports change at different rates and the time windows differ from
weeks to months.
}
\label{fig:spaghettis}
\end{figure} 

\begin{figure} 
\includegraphics[width=0.97\textwidth]{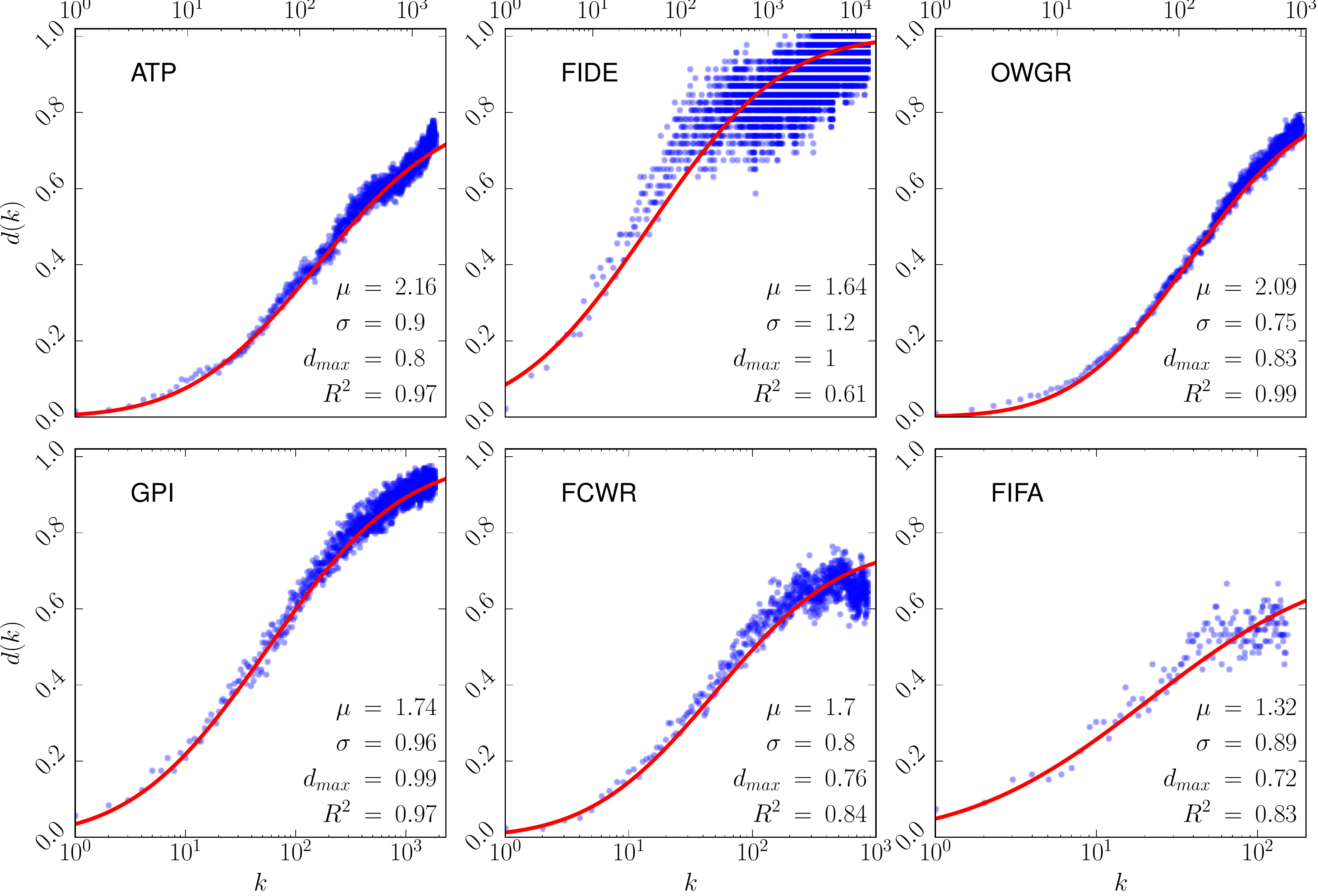}
\caption{\csentence{Rank diversity of sports and games.} 
Plot showing the rank diversity $d(k)$ for all datasets ({\color{blue} blue}
dots), as well as the fit $\Phi$ ({\color{red} red} lines). Values of $\mu$,
$\sigma$, maximum rank diversity $d_{max}$, and $R^2$ are also shown.
}
\label{fig:diversity}
\end{figure} 
\begin{figure} 
\includegraphics[width=\textwidth]{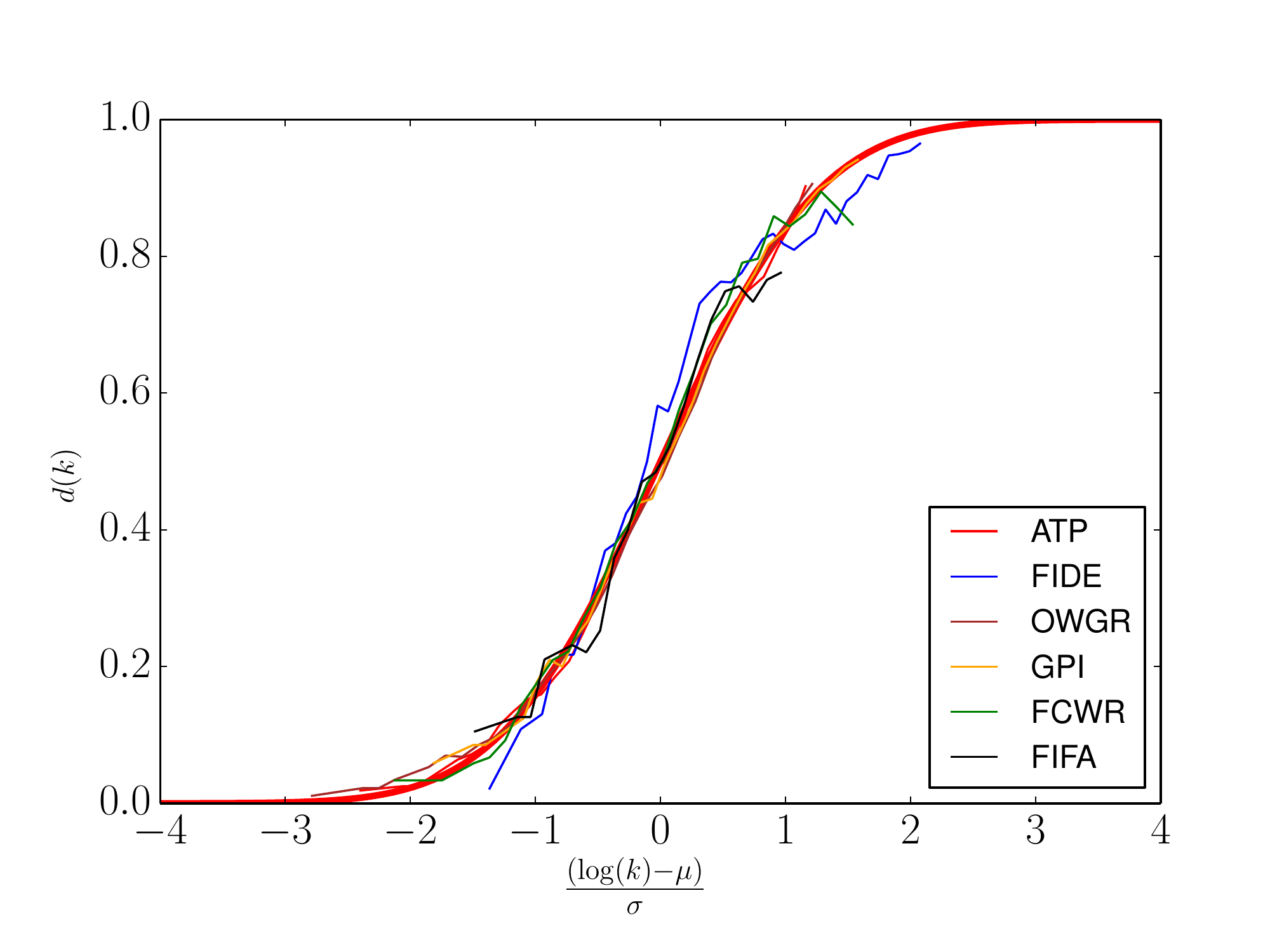}
\caption{\csentence{Similarity in the normalized rank diversity across sports
and games.} Plot showing a comparison of the rank diversity $d(k)$ for all
activities considered. With the values of $\mu$ and $\sigma$ obtained from the
fit $\Phi$, we have normalised $d(k)$ by rank $k$ with
$\frac{\log(k)-\mu}{\sigma}$. As reference we include the basic form of
\eref{eq:Phi} (thick {\color{red} red} line), indicating that all activities
have the same functional shape of rank diversity.
}
\label{fig:normalized}
\end{figure} 

\begin{figure} 
\includegraphics[width=0.97\textwidth]{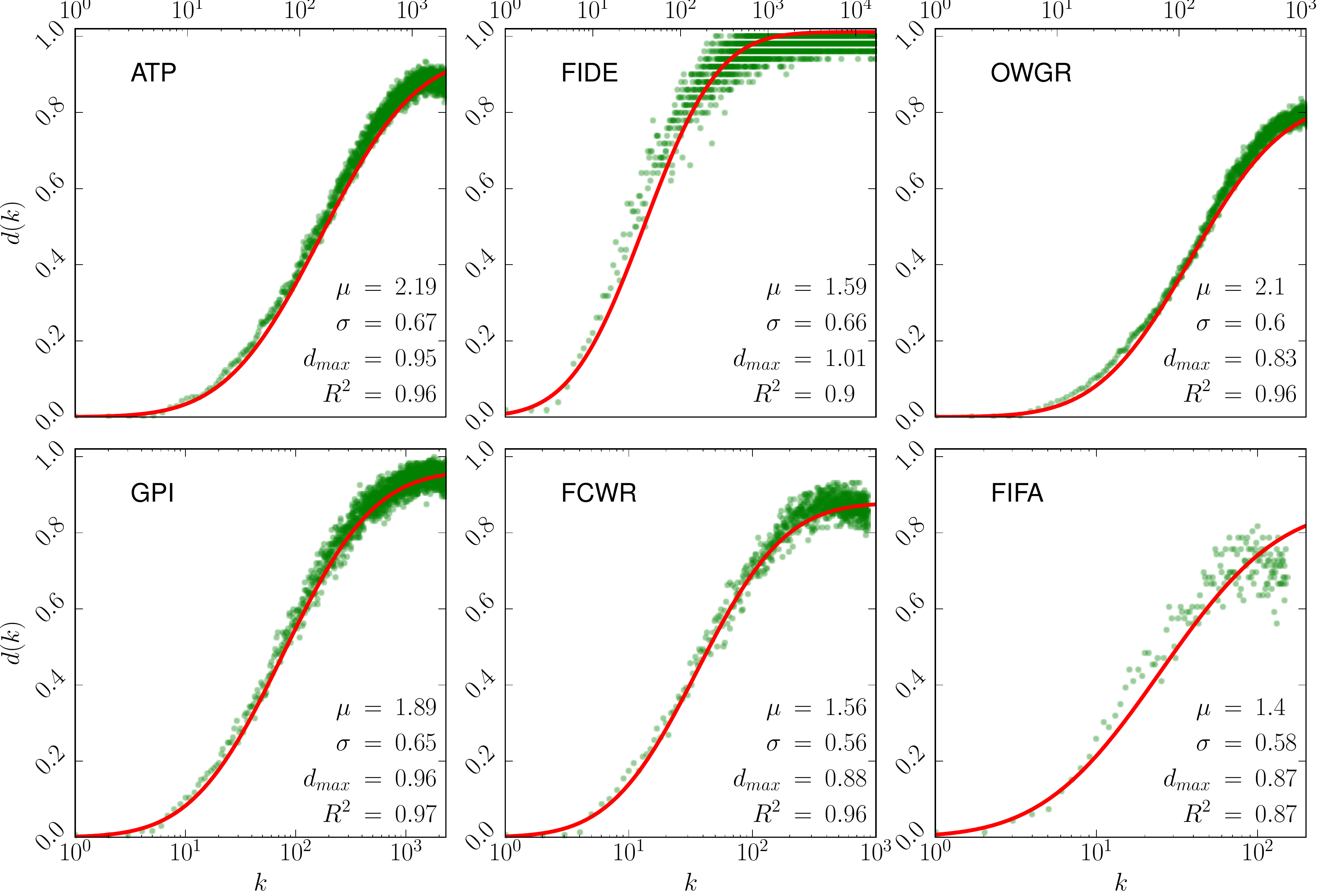}
\caption{\csentence{Simulated rank diversity.} 
Plot showing the rank diversity $d(k)$ coming from our random walk model, with
$\hat \sigma$ values adjusted from empirical data. We include the values of
$\mu$, $\sigma$, maximum rank diversity $d_{max}$, and $R^2$, as well as the
basic form of \eref{eq:Phi} (thick {\color{red} red} line).
}
\label{fig:diversity:simulated}
\end{figure} 

\clearpage

\section*{Tables} 

\begin{table*}[ht] 
\caption{ {\bf Summary of ranking data for each sport and game considered in
this study.} Table listing the main properties of the ranking data used here
(including data source, time period, ranking resolution, and number of players
or teams). In order to have a homogeneous ranking resolution for each sport and
game, we disregard some data for the ATP, FIDE, and GPI datasets.}
\label{tab:ranking_data}
\begin{tabular}{ p{1.8cm} p{3.1cm} p{2cm} p{2.0cm} p{2.1cm} }
\toprule
 Sport/game & Data source & Time period & Time resolution & \# players/teams \\
\midrule
\noalign{\smallskip}
 Tennis players \newline (male) & Association of Tennis \newline Professionals (ATP)~\cite{atpWorldTour} & May 5 2003 -- \newline Dic 27 2010 & Weekly & 1600\\
\noalign{\smallskip}
 Chess players \newline (male) & F\'ed\'eration Internationale des \'Echecs (FIDE)~\cite{worldChessFed} & Jul 2012 -- \newline Apr 2016 & Monthly & 13500\\
\noalign{\smallskip}
 Golf players & Official World Golf Ranking (OWGR)~\cite{officialWorldGolf} & Sept 10 2000 -- Apr 19 2015 & Weekly & 1000\\
\noalign{\smallskip}
 Poker players & Global Poker Index \newline (GPI)~\cite{globalPokerInd} & Jul 25 2012 -- Jun 10 2015 & Weekly & 1799\\
\noalign{\smallskip}
 Football teams & Football Club World \newline Ranking (FCWR)~\cite{footballClubWorld} & Feb 1 2012 -- Dec 29 2014 & Weekly & 850\\
\noalign{\smallskip}
 National \newline football teams & F\'ed\'eration Internationale \newline de Football Association \newline (FIFA)~\cite{fifa} & Jul 2010 -- \newline Dec 2015 & Monthly & 150\\
\bottomrule
\end{tabular}
\end{table*} 

\begin{table} 
\caption{
{\bf Parameter values for fitting process between sports data and ranking models.}
Table listing parameter values for all models in \esref{eq:m:general}{eq:double:zipf}, obtained in the fitting process with empirical data. These values correspond to the model curves in \fref{fig:rank} (model $m_5$ not shown there). 
}
\label{tab:fitting_rank_data}
\begin{center}
\begin{tabular}{|l|c| c | c | c |c | c | c |c|}\hline 
 & \multicolumn{2}{|c|}{model $m_1$}  & \multicolumn{3}{|c|}{model $m_2$}  & \multicolumn{3}{|c|}{model $m_3$}\\ \hline
&  $\log \mcN$ & $a$ & $\log \mcN$ & $a$ & $b$ & $\log \mcN$ & $a$ & $d$ \\ \hline
ATP  & 4.51& 1.02&   4.16&  0.675               &     2.36$\times 10^{-3}$&    -1.15	 & 0.860&                1.64		\\
FIDE & 3.48& 0.0355& 3.46&  2.30$\times 10^{-2}$&     2.12$\times 10^{-6}$&     2.58	 & 2.68$\times 10^{-2}$& 0.169	\\
OWGR & 1.37& 0.723&  1.05&  0.377               &     2.76$\times 10^{-3}$&    -5.72	 & 0.466&                2.15		\\
GPI  & 3.75& 0.234&  3.66&  0.144               &     6.63$\times 10^{-4}$&     2.54	 & 0.193&                0.358	\\
FCWR & 4.52& 0.529&  4.24&  0.218               &     3.06$\times 10^{-3}$&     2.19	 & 0.341&                0.732	\\
FIFA & 3.43& 0.444&  3.18&  2.38$\times 10^{-2}$&     1.63$\times 10^{-2}$&     0.497    & 0.205&                1.19		\\ \hline
\end{tabular}\\
\end{center}
\begin{center}
\begin{tabular}{|l|c| c | c | c |c | c | c |c|}\hline 
 & \multicolumn{4}{|c|}{model $m_4$}  & \multicolumn{4}{|c|}{model $m_5$} \\ \hline
&  $\log \mcN$ & $a$ & $b$ & $d$ & $\log \mcN$ & $a$ &$a'$ &  $k_c$ \\ \hline
ATP  & 4.16	&0.675 & 2.36$\times 10^{-3}$ & 9.02$\times 10^{-9}$& 4.23 & 0.770 & 2.88   & 3.73$\times 10^{2}$	 \\
FIDE & 3.02	&0.0242 &1.19$\times 10^{-6}$ & 8.52$\times 10^{-2}$& 3.47 & 0.0263 & 0.177 & 2.9641$\times 10^{4}$ \\
OWGR & -0.323	&0.389 & 2.24$\times 10^{-3}$ & 0.435               & 1.11 & 0.465 & 2.03   & 2.54$\times 10^{2}$	 \\
GPI  & 3.66	&0.144 & 6.63$\times 10^{-4}$ & 3.64$\times 10^{-9}$& 3.65 & 0.133 & 0.437  & 1.08$\times 10^{2}$	 \\
FCWR & 2.93	&0.269 & 1.39$\times 10^{-3}$ & 0.458               & 4.35 & 0.371 & 3.47   & 4.26$\times 10^{2}$ \\
FIFA & 0.734	&0.176 & 1.95$\times 10^{-3}$ & 1.08	            & 3.32 & 0.318 & 2.88   & 81.7	 \\ \hline
\end{tabular}
\end{center}
\end{table} 

\begin{table} 
\caption{{\bf Goodness of fit measures}. 
Table listing values of $R^{2}$, $D$, and $p$
for the fitting process between the six sports and five theoretical rank
distributions. Higher $R^{2}$ and lower $D$ imply better fits. The best fits
for each sport are shown in \textbf{bold}.}
\begin{tabular}{|l c| c | c | c |c | c |} 
 \hline
     & & $m_1$ & $m_2$ & $m_3$ & $m_4$ & $m_5$ \\
 \hline
 \hline
    &  $R^2$ & 0.165 & 0.987  &  0.858 & \textbf{0.988} & 0.962 \\
ATP &  $D$   & 0.180 & 0.079  &  0.107 & \textbf{0.0724}&  0.219 \\
    &  $p$   & 0.01  & 0.23  &  0 &     \textbf{0.73} & 0.0 \\
 \hline
     &  $R^2$   & 0.086 & 0.913 & 0.967 & \textbf{0.995} & 0.843 \\
FIDE &  $D$     & 0.627 & 0.12  & 0.051 & \textbf{0.028} & 0.359 \\
     &  $p$     & 0.0   & 0.0   & 0     & 0.0   & 0.02 \\
 \hline
     &  $R^2$ & 0.661 & \textbf{0.993} & 0.978 & \textbf{0.993} & 0.972 \\
OWGR &  $D$   & 0.478 & \textbf{0.021} & 0.082 & 0.029 & 0.155 \\
     &  $p$   & 0.0   & \textbf{0.99}  & 0.0   & 0.23 & 0.0 \\
 \hline
      &  $R^2$  & 0.801 & 0.972 & 0.934 & 0.979 & \textbf{0.993} \\
GPI   &  $D$    & 0.514 & 0.204 & 0.154 & 0.214 & \textbf{0.139} \\
      &  $p$    & 0.0   & 0.0   & 0.0   & 0.0   & 0.0 \\
 \hline
     &  $R^2$ & 0.802 & 0.988 & 0.991 & \textbf{0.997} & 0.955 \\
FCWR &  $D$   & 0.282 & 0.127 & 0.059 & \textbf{0.055} & 0.165\\
     &  $p$   & 0.00 & 0.0   & 0.0   & 0.0   & 0.0 \\
 \hline
     &  $R^2$ & 0.776 & 0.972 & 0.996 & \textbf{0.997} & 0.953 \\
FIFA &  $D$   & 0.392 & 0.133 & \textbf{0.038} & 0.047 & 0.144   \\ 
     &  $p$   & 0.0   & 0.0   & \textbf{0.58} & 0.18 & 0.07 \\\hline
\end{tabular}
\label{table:goodnes:of:fits}
\end{table} 
\end{backmatter}
\end{document}